\documentclass[twocolumn,showpacs,preprintnumbers,amsmath,amssymb,prl]{revtex4}
\usepackage{graphicx}% Include figure files
\usepackage{dcolumn}% Align table columns on decimal point
\usepackage{bm}% bold math
\usepackage{multirow}
\usepackage{color}

\newcommand{\be}{\begin{equation}}
\newcommand{\ee}{\end{equation}}

\newcommand{\bea}{\begin{eqnarray}}
\newcommand{\eea}{\end{eqnarray}}

\begin{document}

\title{Coupled-channels calculations of nonelastic cross sections using a density-functional structure model}

\author{G. P. A. Nobre}
\author{F. S. Dietrich}
\author{J. E. Escher}
\author{I. J. Thompson}
\affiliation{Lawrence Livermore National Laboratory, P. O. Box 808, L-414, Livermore, CA 94551, USA}
\author{M. Dupuis}
\affiliation{CEA, DAM, DIF,  F-91297 Arpajon, France}
\author{J. Terasaki}
\author{J. Engel}
\affiliation{Department of Physics and Astronomy, University of North Carolina, Chapel Hill, North Carolina 27599-3255}

\begin{abstract}
A microscopic calculation of the reaction cross-section for nucleon-nucleus scattering has been performed by explicitly coupling the elastic channel  to all particle-hole (p-h) excitation states in the target and to all one-nucleon pickup channels. 
The p-h states may be regarded as {\em doorway states} through which the flux flows to more complicated configurations, and 
subsequently to long-lived compound nucleus resonances. 
Target excitations for $^{40,48}$Ca, $^{58}$Ni, $^{90}$Zr and $^{144}$Sm were described in a QRPA framework using a Skyrme
functional.  
Reaction cross sections calculated in this approach
were compared to predictions of a fitted optical potential and to experimental data, reaching very good agreement. 
Couplings between inelastic states were found to be negligible, 
while the couplings to pickup channels contribute significantly.
For the first time observed reaction cross-sections are completely accounted for by explicit channel coupling, 
for incident energies between 10 and 40 MeV.
\end{abstract}
\pacs{24.10.Eq, 25.40.-h, 24.50.+g}
\date{\today}

\maketitle

A quantitative description of nucleon-nucleus reactions is crucial for a broad variety of applications, including astrophysics, nuclear energy, 
radiobiology and space science \cite{FrontiersNuclSci2007,Townsend2002ASR30}. A fully microscopic description of such reactions is quite complex and resource-consuming, as one needs to consider not only the desired outcome
in an exit channel, but also the interference and competition with all other possible outcomes. 
A successful account of elastic nucleon-nucleus scattering, for example, has to include the effects from the excitation of non-elastic 
degrees of freedom, such as collective and particle-hole (p-h) excitations, transfer reactions, etc. Formally, these non-elastic
effects can be accounted for by the projection-operator approach of Feshbach \cite{Feshbach1958ARNS8}. 
The picture that emerges is one in which flux is removed from the elastic channel by couplings to the non-elastic degrees of freedom. 
An optical potential can therefore be defined \cite{Feshbach1958ARNS8,Brown1959RMP31} as the effective interaction in a 
single-channel calculation that contains
the effects of all the other processes that occur during collisions between nuclei.
Optical potentials play a very important role in the description of nuclear reactions. They are extensively used to describe the interactions of
projectile and target in the entrance channel, and the interaction of ejectile and residual nuclei after the reaction; they are crucial
ingredients in direct-reaction as well as statistical (Hauser-Feshbach) calculations.

Most widely used are phenomenological optical potentials fitted to reproduce experimental data sets. 
They have been extremely successful for many applications involving nuclei in the range of the fits. 
Unfortunately, such adjustable potentials make strong assumptions about locality and momentum dependence that are probably not justified. In addition, for nuclei lying outside the range of the fits, such as the nuclei produced at rare-isotope facilities, in the $r$-process, and in advanced reactor applications, this can lead to unquantifiable uncertainties. To achieve a better understanding of nuclear reactions and structure it is important to calculate optical potentials by first-principle methods.

Within microscopic reaction theory, an optical potential is comprised of two components. 
The first is a real bare potential, the diagonal potential within the elastic channel, which is generally obtained by folding the nucleon distributions of  both nuclei with a nucleon-nucleon effective interaction. 
The second is a complex dynamic polarization potential which arises from couplings to inelastic states. 
The resulting optical potential is composed of an imaginary potential and a real part usually slightly different from the bare potential. The former gives rise to absorption of flux from the elastic channel to the other reaction channels, and is hence directly connected with observed reaction cross-sections.

Several attempts have been made to generate optical potentials from microscopic approaches. Some have used the so-called nuclear matter approach \cite{Jeukenne1977PRC16}, which provides accurate results
at nucleon energies $\gtrsim$ 50 MeV \cite{Barbieri2005PRC72}. Recently, new methods based on self-energy theory have  been implemented \cite{Amos2000ANP25}, and new calculations, which
combine a nuclear matter approach and  Hartree-Fock-Bogoliubov (HFB) mean field structure model, provide encouraging results for neutron scattering below 15 MeV \cite{Pilipenko2010PRC81}. Earlier attempts used the nuclear structure approach, which is more suitable at energies below ~50 MeV \cite{Bouyssy1981NPA371}, and calculated second-order diagrams using particle-hole propagators in the RPA approximation \cite{VinhMau1976NPA257,Bernard1979NPA327,Osterfeld1981PRC23,Bouyssy1981NPA371}. However, these were not able to fully explain observed absorption: e.\ g., in Ref.\ \cite{Bernard1979NPA327}, the couplings could account only for $\approx$ 44\% of the nucleon-nucleus absorption and, in Ref.\ \cite{Osterfeld1981PRC23}, only for $\approx$ 71\% including charge exchange.

\begin{figure}[t]
 \begin{center}
  \includegraphics[trim = 5mm 17mm 23mm 30mm, clip,scale=0.44,angle=0.0]{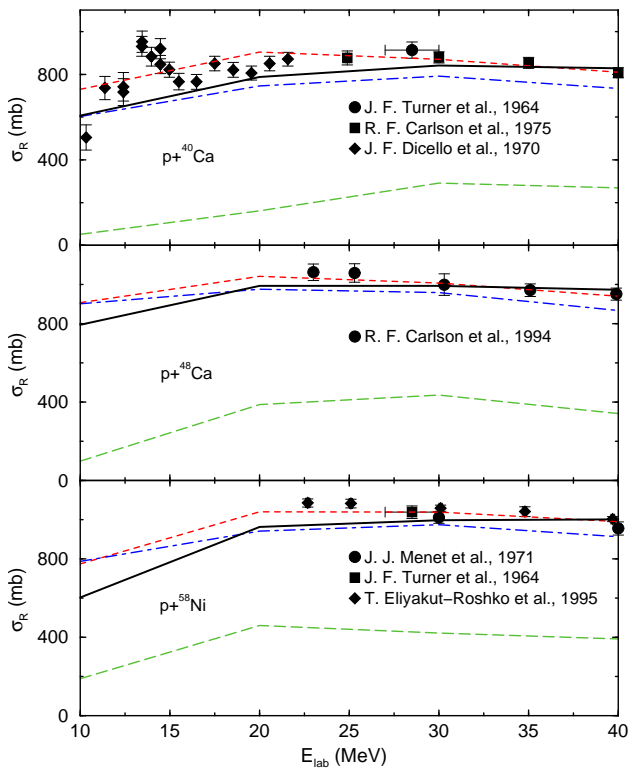} \\
 \end{center}
 \vspace{-5mm}
 \caption{(color online). Total reaction cross-section as a function of the incident energy for p + $^{40}$Ca,  p + $^{48}$Ca and  p + $^{58}$Ni. The results are shown for couplings to the inelastic states lying below 30 MeV (dashed lines), to the inelastic and transfer  channels (dash-dotted lines) and to the inelastic and transfer channels with non-orthogonality corrections (solid lines). The Koning-Delaroche  \cite{Koning2003NPA713} optical model calculations are shown as short-dashed lines. The lines serve as guidance to the eye as calculations were performed only for $E_{\mathrm{lab}}$ = 10, 20, 30 and 40 MeV. Data from Refs.\ \cite{Turner1964NP58,Carlson1975PRC12,Dicello1970PRC2,Carlson1994PRC49,Menet1971PRC4,Eliyakut-Roshko1995PRC51}.}
 \label{Fig:SigRxElabData}
 \vspace{-3.0mm}
\end{figure}

In this Letter, we report on the first major step towards achieving a complete microscopic calculation of the reaction cross sections for both
neutron and proton induced reactions on a variety of medium-mass targets.  A summary of our results 
for protons on $^{40,48}$Ca and  $^{58}$Ni  is given in Figure \ref{Fig:SigRxElabData}, where our best
predictions are the solid lines. The other curves will be described below, where we discuss our method in more detail.

Our calculations aim to obtain the reaction cross sections of all the open channels that can be reached in one step from elastic scattering.
We then use a {\em  doorway approximation}, which takes the total flux leaving the elastic channel to all possible first-order channels to be independent of what happens afterwards:  a nucleon later might escape as a free nucleon, the flux might equilibriate to compound-nuclear resonances, etc.
 
To generate sets of excited states, we use RPA and QRPA structure models for finite nuclei, which start
from HFB  structure models based on energy-density functionals. For each excited state, we
calculate the one-body transition density and corresponding transition potential by the methods of \cite{DupuisPhDthesis,Terasaki2005PRC71}, 
which we use within large coupled-channels calculations. In addition to inelastic excitations, we also include couplings to pickup channels. 

To obtain the initially  
occupied proton and neutron levels in a nucleus, we use the Skryme energy-density functional SLy4 \cite[Table 1]{Chabanat1998NPA635}, a 
parametrization designed to  describe systems with arbitrary neutron excess, from stable to neutron matter, by improving isotopic properties, which overcomes deficiencies of other interactions away from the stability line.
A HFB calculation gives  the particle and hole levels of a given nucleus and fixes the p-h basis states for generating excited states within the framework of (Q)RPA,  thus accounting for correlations caused by the residual interactions within the target.

Our scattering effective nucleon-nucleon interaction is of Gaussian shape, with parameters matched to  the volume integral and r.m.s.\ radius of the M3Y interaction at 40 MeV; it  includes a knock-on exchange correction \cite{Love}. % to avoid explicit calculation of exchange terms.
In momentum space,  the central effective interaction is
$
v^{T}(q)= V_{0}^{T} (\pi /\mu_{T}^2)^{3/2} e^{-q^{2}/(2\mu_{T})^{2}},
%\label{Eq:GaussVq}
$
%A
with $V_0^{0} = -24.1921$ MeV and $\mu_{0} = 0.7180$ fm$^{-1}$ for the isoscalar part of the interaction and $V_0^{1} = 11.3221$ MeV and $\mu_{1} = 0.7036$ fm$^{-1}$ for the isovector component.
We do not include any imaginary part in this effective interaction, as our aim is to include all non-elastic excitations explicitly in our model.
We convolute $ v^{T}(q)$ with the transition densities to generate the configuration-space transition potentials.
The bare potential in the elastic channel is the single-folded potential using the ground-state density from the HFB calculation. For simplicity, this potential was also used for all excited states.

To explore the relative importance of the various contributions to the reaction cross section, we carried out a series of calculations:

(1) Inelastic coupled-channels calculations will be shown for reactions involving protons and neutrons scattered by the nuclei $^{40}$Ca, $^{48}$Ca, $^{58}$Ni, $^{90}$Zr and $^{144}$Sm, coupling the ground state to all levels with excitation energy ($E^{*}$) lying below some limit,  according to the QRPA model. The QRPA states above the particle emission threshold are used to approximate exact scattering waves. Recent studies  have shown that such wave functions contain large density distributions outside the nuclear radius \cite{Terasaki2007PRC76}. When used in reaction calculations they accurately represent the continuum \cite{Moro2007PRC75}. Thus, processes containing one nucleon in the continuum (plus the inelastically scattered projectile) are included in our model.

(2) Additional couplings \emph{between} excited states were considered as predicted by the  RPA model.  
These couplings, however, were found to be negligible for  scattering energies above 10 MeV, allowing us to disregard them in the subsequent calculations.

(3) Consideration of a finite-range  interaction in a HFB description of the target structure. For reactions of nucleons scattered by $^{90}$Zr, the reaction cross section results using the QRPA model with the SLy4 force were found to be practically equivalent to the results
found using RPA states and transitions with the Gogny D1S force \cite{Decharge1980PRC21}. This was observed despite the proton pairing gap of 1.2 MeV of  $^{90}$Zr. 

(4) Inclusion of one-nucleon pick-up channels: pick-up channels  play an important role in nucleon-nucleus scattering \cite{Mackintosh-Keeley,Coulter1977NPA293}. 
Coupled reaction channels (CRC) calculations were performed, including all  the channels for the formation of a deuteron, picking up the appropriate nucleon from occupied levels in the target.
For transfers, we approximate the HFB target states by bound single-particle states in a Woods-Saxon potential,  with the radii fitted to reproduce the volume radii and Fermi energy obtained by the HFB calculations.
The volume diffuseness and spin-orbit parameters were taken from Koning-Delaroche optical potentials \cite{Koning2003NPA713} 
at $E_{\rm lab}=0$, with spin-orbit radii adjusted by the same factor used to fit the volume part to HFB radii.
To overcome numerical limitations, we coupled explicitly only to the transfer channels, incorporating all inelastic couplings in the inelastic optical potential already calculated in (1).

CRC calculations require, in addition to the scattering potentials in the incoming channel, a scattering potential between the deuteron and the remaining target.
We adopted the Johnson-Soper \cite{Johnson1970PRC1} prescription as it includes the effects of deuteron breakup in adiabatic (sudden) approximation. 
In this prescription, the deuteron potential is the sum of the individual  neutron and proton potentials with the target. 
For the real parts we used the diagonal transition potentials  of the corresponding nucleon-nucleus reaction and,
for the imaginary parts, the sum of the  imaginary parts of the Koning-Delaroche  \cite{Koning2003NPA713}  optical potential for protons and neutrons. That is, fitted parameters are used in the imaginary part of the deuteron potential, while we leave for future work to calculate deuteron and nuclear potentials self-consistently.

To assess the success of our large-scale coupled-channels approach, we compare the calculated reaction cross section to that obtained by one of the best available phenomenological optical potentials, henceforth referred to as $\sigma_{\mathrm{R}}^{\mathrm{OM}}$  \cite{Koning2003NPA713}.

\begin{figure}[t]
 \begin{center}
  \includegraphics[trim = 18mm 22mm 17mm 35mm, clip,scale=0.34,angle=0.0]{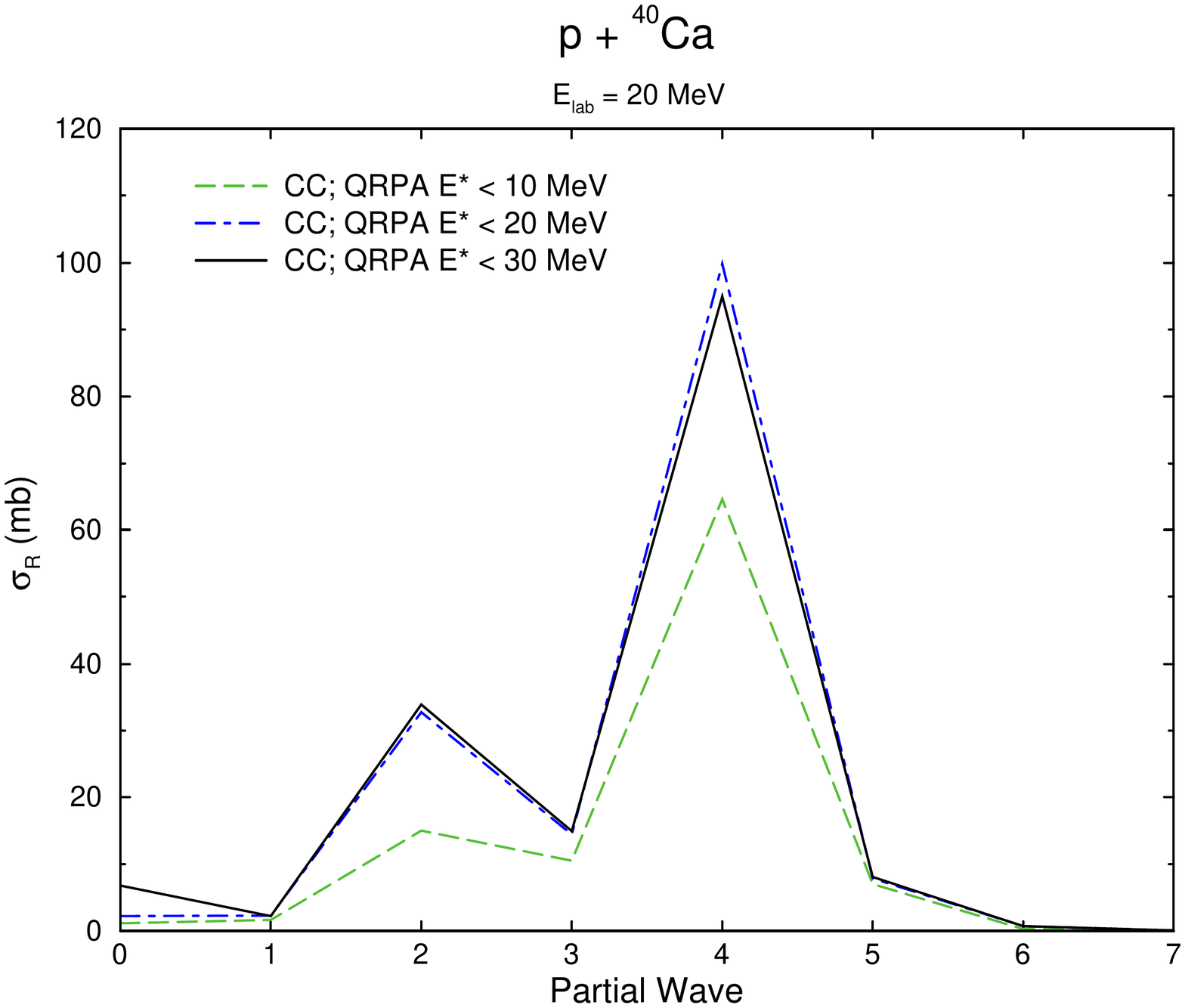} \\
 \end{center}
 \vspace{-5mm}
 \caption{(color online). Reaction cross-section as a function of the partial wave for the reaction p + $^{40}$Ca  at $E_{\rm lab}$ = 20 MeV.}
 \label{Fig:pCa40SigRxJInel}
\end{figure}

\begin{figure}[t]
 \begin{center}
  \includegraphics[trim = 18mm 22mm 17mm 35mm, clip,scale=0.34,angle=0.0]{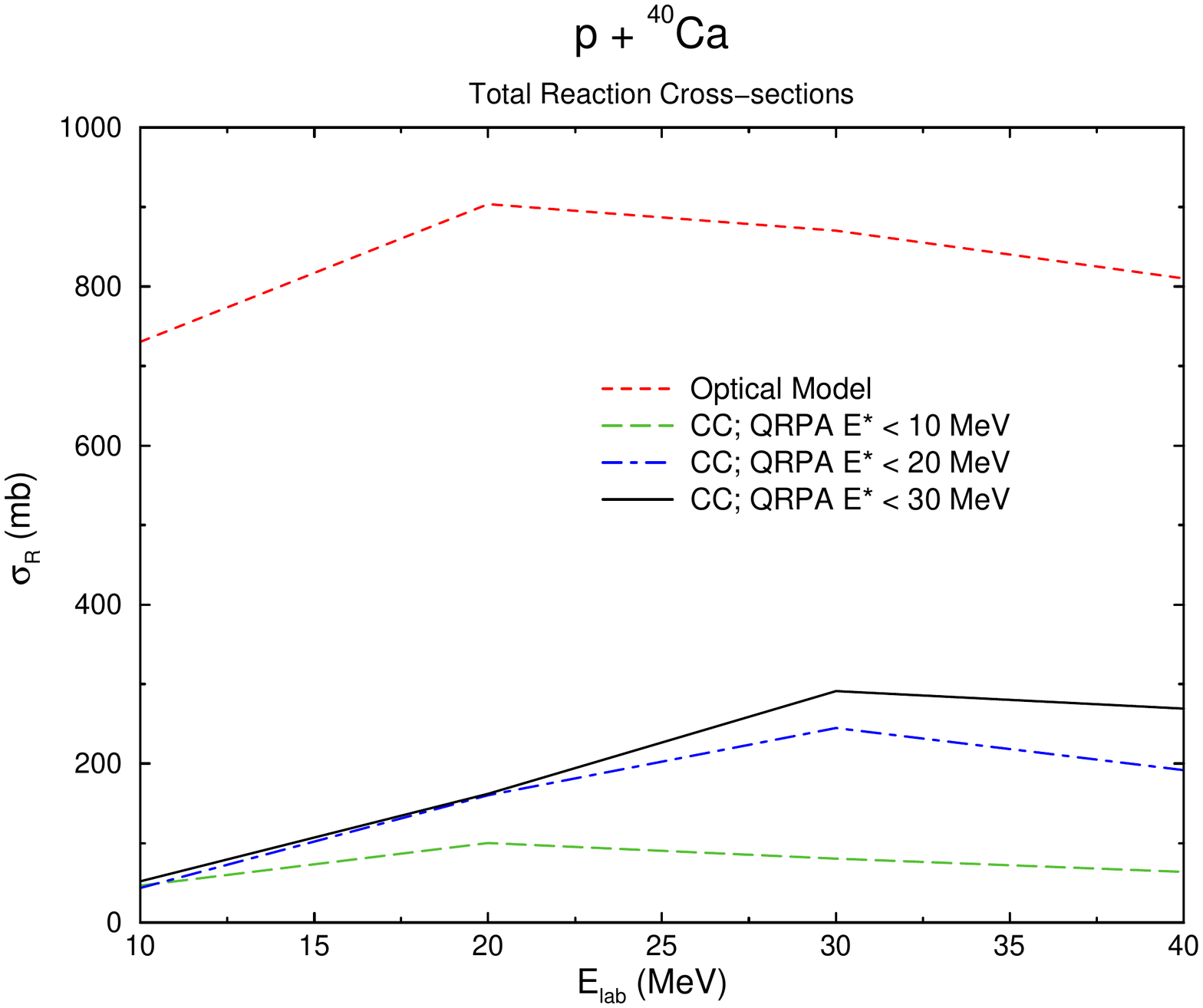} \\
 \end{center}
 \vspace{-5mm}
 \caption{(color online). Total reaction cross-section as a function of the incident energy for the reaction p + $^{40}$Ca, for the different inelastic calculations. The short-dashed line shows the results using the Koning-Delaroche  \cite{Koning2003NPA713} optical potential. The lines serve as guidance to the eye as calculations were performed only for $E_{\mathrm{lab}}$ = 10, 20, 30 and 40 MeV.}
 \label{Fig:pCa40SigRxElab}
  \vspace{-3mm}
\end{figure}

%%\section{Results}

%To determine when all relevant inelastic channels are being explicitly coupled, calculations considering all the excited states below 10, 20 and 30 MeV were performed. 
We examined the convergence with respect to maximum excitation energy, and found that convergence of the inelastic calculations requires coupling of  all excited levels below the scattering energy (i.e.\ all open channels).
This behavior is observed for each partial wave as well as at all energies, as is illustrated for p + $^{40}$Ca in Figs.\ \ref{Fig:pCa40SigRxJInel} and \ref{Fig:pCa40SigRxElab}, respectively.

Although  the reaction cross section increases with the number of coupled states, to the limit where all open channels are coupled, Figure \ref{Fig:pCa40SigRxElab} shows that these inelastic couplings account only for a small fraction ($\approx$ 23\%  at $E_{\rm lab}$ = 30 MeV) of $\sigma_{\mathrm{R}}^{\mathrm{OM}}$. However, after including couplings to the pickup channels through the CRC calculations, a large increment is found, approaching $\sigma_{\mathrm{R}}^{\mathrm{OM}}$ and the experimental data, as can be seen in Figure \ref{Fig:SigRxElabData}.
 An even better agreement can be obtained after we include the non-orthogonality terms \cite[p.\ 226]{ThompsonBookNonOrthogonality} in the CRC calculations, also shown in Figure \ref{Fig:SigRxElabData}. This correction arises because at small radii the deuteron bound state is not orthogonal to bound states occupied in the target.

This work focuses on reaction cross sections, which test the modulus of the S-matrix elements. Additional insights can be gained, e.\ g., from elastic angular distributions. Preliminary calculations of these give reasonable agreement with measured cross sections.

\begin{figure}[t]
 \begin{center}
  \includegraphics[trim = 4mm 17mm 22mm 46mm, clip,scale=0.41,angle=0.0]{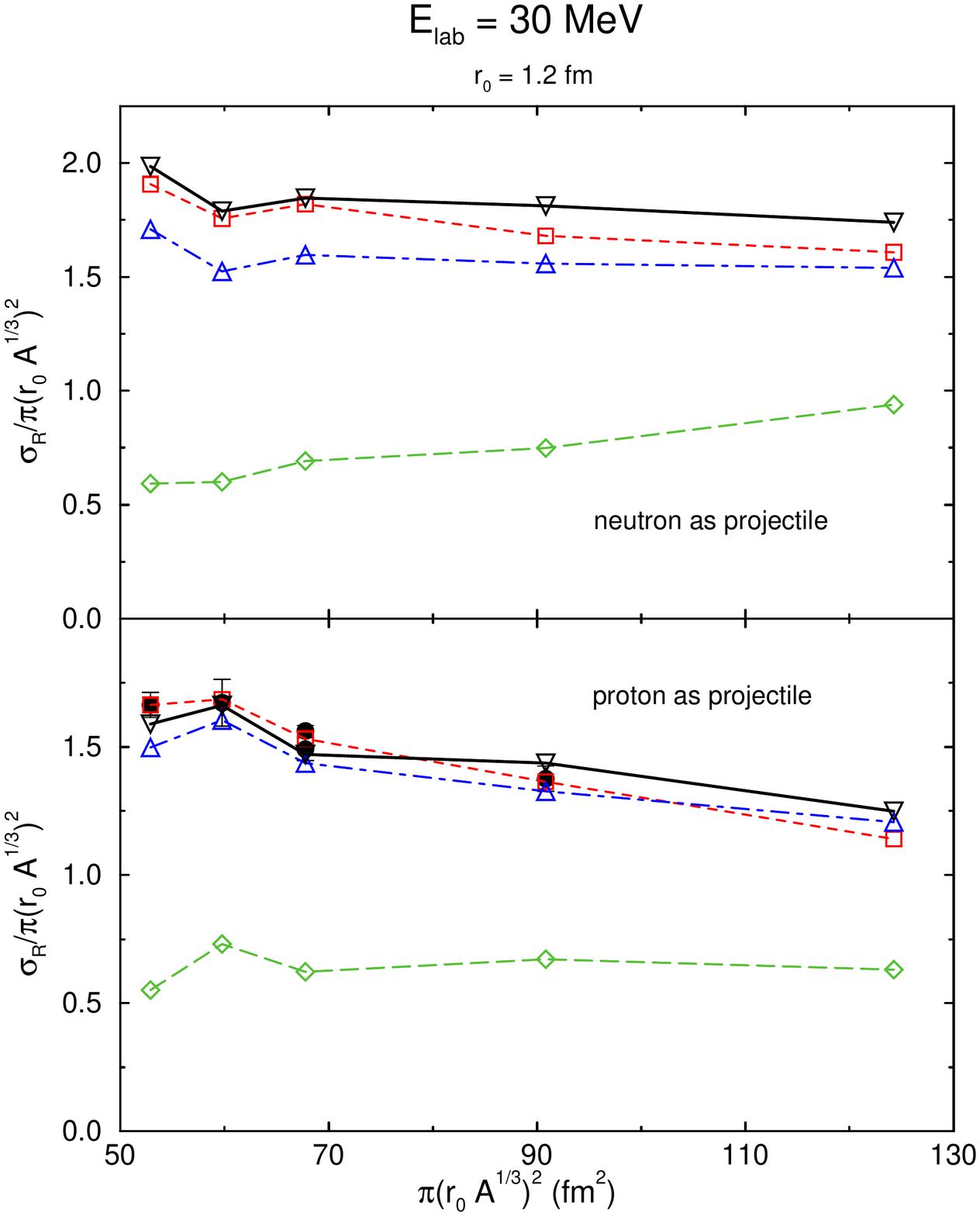} \\
 \end{center}
 \vspace{-5mm}
 \caption{(color online). Normalized total reaction cross-section at $E_{\rm lab}$ = 30 MeV  as a function of the area of the targets $^{40,48}$Ca, $^{58}$Ni, $^{90}$Zr and $^{144}$Sm, with $r_{0}$ = 1.2 fm. The lines have the same meanings as in Figure \ref{Fig:SigRxElabData} and the open symbols are the calculations for the different target nuclei studied. Filled symbols are experimental data from Refs. \cite{Carlson1975PRC12,Carlson1994PRC49,Menet1971PRC4,Eliyakut-Roshko1995PRC51}}
 \label{Fig:allnucleiat30MeV}
 \vspace{-3.0mm}
\end{figure}

In Figure \ref{Fig:allnucleiat30MeV} we present the reaction cross sections obtained for nucleons scattered by the nuclei $^{40,48}$Ca, $^{58}$Ni, $^{90}$Zr and $^{144}$Sm at an incident energy of 30 MeV, as a function of the area  of the target. The absorption is shown relative to the reaction cross-section of a black sphere, which is approximately the geometrical area of the target. It can be seen again that, despite the important contribution of all inelastic couplings to the reaction cross-section, a large amount of absorption is due to the pickup channels and  the corresponding non-orthogonality corrections. Explicitly considering such couplings enabled us to account for practically all of the non-elastic cross-sections in the studied reactions.

%%\section{Conclusion}

In summary, we have calculated the reaction cross-sections for nucleon induced reactions  on nuclei    $^{40,48}$Ca, $^{58}$Ni, $^{90}$Zr and $^{144}$Sm, by explicitly calculating the couplings to all the doorway transfer and (Q)RPA inelastic  channels. We found that inelastic convergence is achieved when all open channels are coupled. 
While inelastic couplings account for an important part of the reaction cross section, most contributions come from couplings to the deuteron pickup channel,
in which case the non-orthogonality terms are significant. 
We obtain reaction cross sections that are in good agreement with phenomenological optical model results and experimental data. Such results, using the doorway approximation, are an important milestone. Future work on couplings between different types of non-elastic processes will calculate higher-order corrections.

This work represents the first complete microscopic calculation that uses basic interactions between nucleons  within the nuclei to predict reaction observables for incident energy as low as 10 MeV.
Using state-of-the-art nuclear structure models coupled with large-scale reaction computations allowed the accurate prediction of measurable quantities. This will serve as basis for future fully-consistent \emph{ab initio} developments for a range of nuclei including unstable species.

%\begin{theacknowledgments}
%%\subsection{Acknowledgments}
 This work was performed under the auspices of the U.S. Department of Energy by Lawrence Livermore National Laboratory under Contract DE-AC52-07NA27344, and under SciDAC Contract DE-FC02-07ER41457.
%\end{theacknowledgments}

%\bibliographystyle{plainnat}  %% nome-ano
\bibliographystyle{unsrt}    %% numero
\bibliography{ReactionPRL}

\end{document}